\documentclass[journal]{IEEEtran}

\usepackage{amsfonts}
\usepackage{amsmath}
\usepackage{amssymb}

\usepackage{authblk}

\usepackage{graphicx}
\usepackage{graphics}
\usepackage{multirow}
\usepackage{dblfloatfix}
\usepackage{cite}
\usepackage{xcolor}
\usepackage{soul,color}
\usepackage{subcaption}
\usepackage{soul}

\begin{document}

\title{Deep Extended Feedback Codes}

\author{Anahid~Robert~Safavi${}^{(1)}$, Alberto~G.~Perotti${}^{(1)}$, Branislav~M.~Popovi\'c${}^{(1)}$, \\
		Mahdi~Boloursaz~Mashhadi${}^{(2)}$, Deniz~G\"und\"uz${}^{(2)}$
		\thanks{(1) Radio Transmission Technology Lab, Huawei
		Technologies Sweden AB, Kista 164-94, Sweden}
				\thanks{(2) Information Processing
					and Communications Laboratory, Department of Electrical and Electronic
					Engineering, Imperial College London, London SW7 2BT, U.K.}}

\maketitle

\begin{abstract}
A new deep-neural-network (DNN) based error correction encoder architecture
for channels with feedback, called Deep Extended Feedback (DEF), is presented in this paper.
The encoder in the DEF architecture transmits an information message
followed by a sequence of parity
symbols which are generated based on the message as well as the observations of the past
forward channel outputs sent to the transmitter through a feedback channel.
DEF codes generalize Deepcode~\cite{Kim2018} in several ways:
parity symbols are generated based on forward-channel output observations
over longer time intervals in order to provide better error correction capability;
and high-order modulation formats are deployed in the encoder so as to achieve increased
spectral efficiency.
Performance evaluations show that DEF codes have better performance compared
	to other DNN-based codes for channels with feedback.
\end{abstract}

\IEEEpeerreviewmaketitle

\section{Introduction}

The fifth generation (5G) wireless cellular networks' New Radio (NR) access technology
 has been recently specified by the $3^{\rm rd}$ Generation Partnership
Project (3GPP).
NR already fulfills demanding requirements of throughput, reliability and latency.
However, new use cases stemming from new application domains (such as industrial
automation, vehicular communications or medical applications) call for further
significant enhancements.
For instance, some typical Industrial Internet of Things (IIoT) applications
would need considerably higher reliability and shorter transmission delay compared
to what 5G/NR can provide nowadays.

Error correction coding is a key physical layer functionality for
guaranteeing the required performance levels.
In conventional systems, error correction is accomplished by linear binary codes
such as polar codes~\cite{Arikan2009},
Low Density Parity Check (LDPC) codes~\cite{Frey2000} or turbo codes~\cite{Berrou93}, possibly combined with retransmission mechanisms such as Hybrid Automatic Request (HARQ)~\cite{Lin}.
HARQ performs an initial transmission followed by a variable number of subsequent
incremental redundancy transmissions until the receiver notifies successful decoding
to the transmitter.
Short acknowledgment (ACK) or negative ACK (NACK) messages are sent through a
feedback channel in order to inform the transmitter about decoding success.
By usage of simple ACK/NACK feedback messages,
conventional HARQ practically limits the gains that could potentially be obtained
by an extensive and more efficient use of the feedback channel.
Codes that make full use of feedback potentially achieve improved
performance compared to conventional codes, as predicted in~\cite{Polyaskiy2011}.

Finding good codes for channels with feedback is a notoriously difficult problem.
Several coding methods for channels with feedback have been proposed -- see for
example~\cite{Schalkwijk1966,Horstein1963,Ooi1998,Chance_love2015,Vakilinia2016}.
However, all known solutions either do not approach the performance predicted in~\cite{Polyaskiy2011} or exhibit unaffordable complexity.
Promising progress has been made recently by applying Machine Learning (ML)
methods~\cite{Kim2018}, where both encoder and decoder are implemented as two
separate Deep Neural Networks (DNNs).
The DNNs' coefficients are determined through a joint encoder-decoder training
procedure whereby encoder and decoder influence each other.
In that sense, the chosen \emph{decoder} structure has impact on the resulting
code -- a previously unseen feature.
Known DNN-based feedback codes~\cite{Kim2018} use different recurrent
neural network (NN) architectures -- recurrent NNs (RNNs) and gated
recurrent units (GRUs) are used in~\cite{Kim2018}; long-short term
memory (LSTM) architectures have been mentioned in a preprint
of~\cite{Kim2018} as a potential alternative to RNNs for the encoder.

\begin{figure*}[!t]
	\centering
	\includegraphics[scale=0.5,clip=true,trim=.3cm 3.2cm .1cm 1.5cm]{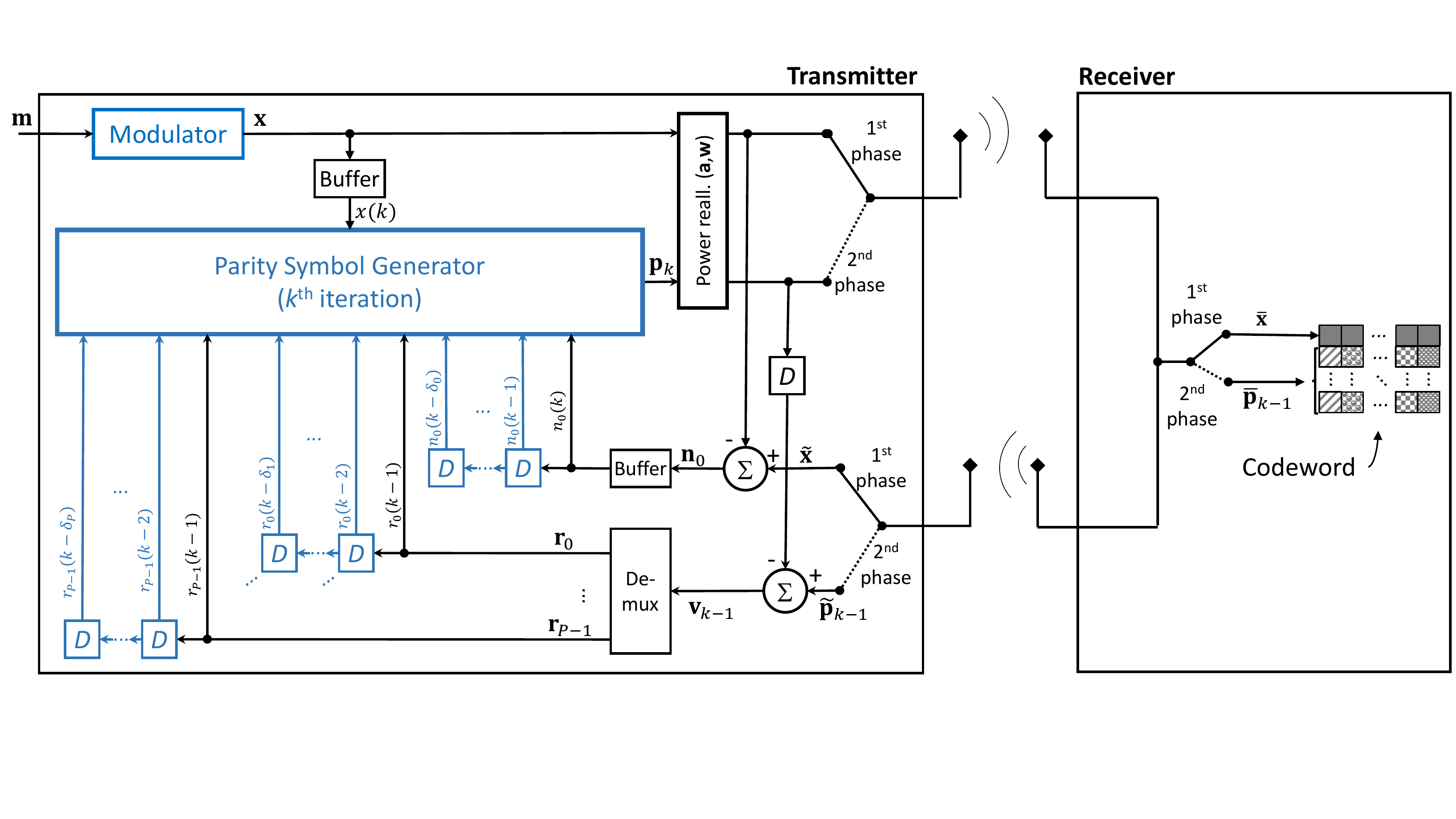}
	\caption{DEF encoder structure.
		Each ``$D$'' block represents a unit-time delay.
		Blue blocks and signals denote new functionalities compared to prior solutions.}
	\label{fig:New procedure}
\end{figure*}

A new DNN-based code for channels with feedback called Deep Extended Feedback
(DEF) code is presented in this paper.
The encoder transmits an information message followed by a sequence of parity
symbols which are generated based on the message and on observations of the past
forward channel outputs obtained through the feedback channel.
Known DNN-based codes for channels with feedback~\cite{Kim2018} 
	compute their parity symbols based on the information message and on the
	most recent information received through the feedback channel.
{The DEF code is based on \emph{feedback extension}, which consists in
extending the encoder input so as to comprise delayed versions of feedback signals.
Thus, the DEF encoder input comprises the most recent feedback {signal} and a set
of past feedback signals within a given time window.
{A similar approach could be used in the decoder to extend its input so as to comprise 
delayed versions of received signals in a given time window.
However, it can be shown that such a generalization of the decoder does not bring any
benefit and therefore it will not be considered in the definition of DEF codes.}
The extended-feedback encoder architecture is combined with different NN architectures
of recurrent type -- RNN, GRU and LSTM.}
The DEF code generalizes Deepcode~\cite{Kim2018}
along several directions.
Its major benefits can be summarized as follows:
\begin{itemize}
  \item \textbf{Improved error correction capability obtained by feedback extension}.
The DEF code generates parity symbols based on feedbacks in a longer time window,
thereby introducing long-range dependencies between parity symbols.
As the above long-range dependencies are a necessary ingredient of all good error
correcting codes, it is expected that feedback extension will bring performance
improvements;
  
  \item \textbf{Higher spectral efficiency obtained by usage of QAM/PAM modulations}.
  The DEF code uses quadrature amplitude modulation (QAM) with arbitrary order,
  thereby potentially achieving higher spectral efficiency;
\end{itemize}

In this work, we initially focus on DEF codes' performance evaluation over channels
with \emph{noiseless feedback}, where the forward-channel output observations are sent
uncorrupted to the encoder.
We also provide preliminary performance evaluations of the DEF code with noisy
feedback.

{\bf{Notation:}}
Lowercase and uppercase letters denote scalar (real or complex) values.
For any pair of positive integers $a$ and $b$ with $a<b$, $[a:b]$
denotes the sequence of integers $[a,a+1,\ldots,b]$, sorted
in increasing order.
Boldface lowercase letters (e.g., $\mathbf{b}$) denote vectors;
unless otherwise specified, all vectors are assumed 
	to be column vectors.
$b(i)$ denotes the $i^{\rm th}$ element of $\mathbf{b}$;
$\mathbf{b}(j:k), j < k,$ denotes the sub-vector that contains the elements of
$\mathbf{b}$ with indices in $[j:k]$.
Boldface uppercase letters like $\mathbf{A}$ denote matrices; $a_{i,j}$ represents
the element of $\mathbf{A}$ in the $i^{\rm th}$ row and $j^{\rm th}$ column.
Notation $f(\mathbf{v})$, where $f$ is a function taking a scalar
input, indicates the vector obtained by applying $f$ to each element
of $\mathbf{v}$.
Hadamard (i.e., element-wise) product is denoted by $\circ$.

\section{Definition of Deep Extended Feedback Code}
\label{GeneralArch}
The Deep Extended Feedback (DEF) code is the set of codewords produced
by the DEF encoder shown in Fig.~\ref{fig:New procedure}.
Blue blocks and signals in Fig.~\ref{fig:New procedure} denote the
new functionalities of the DEF code compared to Deepcode~\cite{Kim2018}
-- \emph{extended feedback} is shown by the unit-time delay blocks labeled ``$D$''
and their corresponding input/ouput signals; QAM/PAM symbols are
produced by the block labeled ``Modulator''.
DEF code and Deepcode operate according to the same
encoding procedure as described afterwards.
The novel DEF code features will be treated in
dedicated subsections.

The encoding procedure consists of two phases.
In the \emph{first phase}, an $L$-bit information message
$\mathbf{m} = (m(0),\ldots,m(L-1))$ is mapped to a sequence of
real symbols $\mathbf{x}=(x(0),\ldots,x(K-1))$, hereafter called
\emph{systematic} symbols.

The modulation sequence $\mathbf{x}$ is transmitted
on the forward channel.
The corresponding sequence $\mathbf{\bar{x}}$ observed by the
receiver is given by
\begin{equation}\label{eq:fwdChannel-Sys}
	\mathbf{\bar{x}}= \mathbf{x} + \mathbf{n}_{0}
\end{equation}
where $\mathbf{n}_{0}$ represents additive white Gaussian noise (AWGN)
and other possible forward-channel impairments.
In the performance evaluations of Section~\ref{Perf},
	$\mathbf{n}_{0}$ is
modeled as a sequence of white Gaussian noise samples.
The receiver stores the observed signal $\mathbf{\bar{x}}$ locally and
immediately echoes it back to the transmitter through the feedback channel.
A corresponding sequence
\begin{equation}
\mathbf{\tilde{x}} = \mathbf{\bar{x}} + \mathbf{g}_{0}
\end{equation}
is obtained at the transmitter, where $\mathbf{g}_{0}$ represents
additive white Gaussian noise and other possible feedback-channel
impairments.

In the \emph{second phase}, for each element  $x(k)$ of $\mathbf{x}$,
	the encoder
computes a corresponding sequence of parity symbols 
\begin{equation}\label{eq:paritysequence}
	\mathbf{p}_{k}=(p_{k}(0),\ldots,p_{k}(P-1)), \quad k=0,\ldots,K-1
\end{equation}
and transmits it through the forward channel.
$P$ is the number of parity symbols that the encoder generates per
systematic symbol.
Thus, the total number of transmitted symbols is $K(1+P)$.
The DEF code rate is defined as the ratio of the message length
$L$ over $K(1+P)$, that is:
\begin{equation}
	R_{\rm DEF} \triangleq \frac{L}{K(1+P)}.
\end{equation}

The receiver observes a set of corresponding parity symbols
sequences $\bar{\mathbf{p}}_{k}, k=0,\ldots,K-1$.
$\bar{\mathbf{p}}_{k}$ can be written as follows:
\begin{equation}\label{eq:RXparitySeq}
	\bar{\mathbf{p}}_{k} = \mathbf{p}_{k} + \mathbf{v}_{k},
\end{equation}
where $\mathbf{v}_{k}=(v_{k}(0),\ldots, v_{k}(P-1))$ represents additive
white Gaussian noise and other  forward channel impairments.
$\bar{\mathbf{p}}_k$ is immediately echoed back to the transmitter through the feedback
channel so as to obtain
\begin{equation}
	\mathbf{\tilde{p}}_{k} = \mathbf{\bar{p}}_{k} + \mathbf{g}_{k},
\end{equation}
where  $\mathbf{g}_k$ represents additive white Gaussian noise and
other feedback channel impairments.

{
The DEF codeword is defined as $\mathbf{z}=(z(0),\ldots, z({(P+1)K-1}))$.
The $j^{\rm th}$ codeword symbol is defined as follows:
\begin{eqnarray}\label{eq:cwelements}
	z(j)\!&\!=\!&\!\left\{\begin{array}{ll}
		w(0) a(j) x(j), & 0\leq j \leq K-1 \\
		w(l+1) a(k) p_{k}(l), & K\!\leq\!j\!\leq (P+1)K-1
	\end{array}\right. \\
	l\!&\!=\!&\!(j-K) \; {\rm mod} \; P, \nonumber \\
	k\!&\!=\!&\!\left\lfloor(j-K) / P \right\rfloor, \nonumber
\end{eqnarray}
where $w(0)$ and $w(l+1), l=0,\ldots, P-1,$ are \emph{codeword}
power levels, $a(k), k=0, \ldots, K-1,$
are \emph{symbol} power levels, $x(j)$ is the $j^{\rm th}$ systematic symbol, 
and $p_{k}(l)$ is the $l^{\rm th}$ symbol of the
$k^{\rm th}$ parity sequence \eqref{eq:paritysequence}.
Codeword power levels reallocate the power among codeword symbols as follows:
the systematic symbols are scaled by $w(0)$;
the $1^{\rm st}$ parity symbol of each parity sequence is scaled by $w(1)$,
the $2^{\rm nd}$ parity symbol of each parity sequence is scaled by $w(2)$, etc.
Symbol power levels reallocate the power among codeword
symbols as follows:
$a(0)$ scales the amplitude of the $1^{\rm st}$ systematic symbol $x(0)$ and of the symbols
of the $1^{\rm st}$ parity symbol sequence $\mathbf{p}_{0}$,
$a(1)$ scales the amplitude of the $2^{\rm nd}$ systematic symbol $x(1)$ and of the symbols
of the $2^{\rm nd}$ parity symbol sequence $\mathbf{p}_{1}$, $\ldots$
$a(K-1)$ scales the amplitude of the $K^{\rm th}$ systematic symbol $x(K-1)$ and of the symbols 
of the $K^{\rm th}$ parity symbol sequence $\mathbf{p}_{K-1}$.
Power levels $w(l)$ and $a(k)$ are obtained by NN training.
The following constraints preserve the codeword's average power:
\begin{equation}
	\sum_{l=0}^{P}w^2(l) = 1, \;\; \sum_{k=0}^{K-1}a^2(k) = 1.
\end{equation}}

\subsection{QAM/PAM Modulator}
\label{subsec:HOM}
The DEF code modulator maps the $L$-bit information message
$\mathbf{m} = (m(0),\ldots,m(L-1))$ to a sequence of
real symbols $\mathbf{x}=(x(0),\ldots,x(K-1))$, hereafter called
\emph{systematic} symbols.
Each pair of consecutive symbols $(x(2i),x(2i+1)), i=0,\ldots,K/2-1$,
forms a complex QAM symbol
$q(i) = x(2i) + x(2i+1)\sqrt{-1}$, where $q(i)$ is obtained by
mapping $Q$ consecutive bits of $\mathbf{m}$ to $2^{Q}$-QAM.
The above mapping produces $K=2L/Q$ real systematic symbols at
the modulator output.

Examples of QAM/PAM mapping of order $Q=2$ and $Q=4$ are shown in
Tab.~\ref{tab:PQAMorder2} and Tab.~\ref{tab:PQAMorder4}.

\begin{table}[!b]
	\centering
	\begin{tabular}{c|cc}
		$m(2i), m(2i+1)$ & $x(2i)$ & $x(2i+1)$\\
		\hline
		$0, 0$ & 1 &  1 \\
		$0, 1$ & 1 & -1 \\
		$1, 0$ & -1 &  1 \\
		$1, 1$ & -1 & -1
	\end{tabular}
\caption{Example of QAM/PAM mapping of order $Q=2$.}
\label{tab:PQAMorder2}
\end{table}

\begin{table}[!h]
	\centering
	\begin{tabular}{c||cc}
		$m(4i), m(4i+1), m(4i+2), m(4i+3)$ & $x(2i)$ & $x(2i+1)$\\
		\hline
		$0, 0, 0, 0$ &  3 &  3 \\
		$0, 0, 0, 1$ &  3 &  1 \\
		$0, 0, 1, 0$ &  3 & -3 \\
		$0, 0, 1, 1$ &  3 & -1 \\
		$0, 1, 0, 0$ &  1 &  3 \\
		$0, 1, 0, 1$ &  1 &  1 \\
		$0, 1, 1, 0$ & -1 & -3 \\
		$0, 1, 1, 1$ & -1 & -1 \\
		$1, 0, 0, 0$ & -3 &  3 \\
		$1, 0, 0, 1$ & -3 &  1 \\
		$1, 0, 1, 0$ & -3 & -3 \\
		$1, 0, 1, 1$ & -3 & -1 \\
		$1, 1, 0, 0$ & -1 &  3 \\
		$1, 1, 0, 1$ & -1 &  1 \\
		$1, 1, 1, 0$ & -1 & -3 \\
		$1, 1, 1, 1$ & -1 & -1
	\end{tabular}
	\caption{Example of QAM/PAM mapping of order $Q=4$.}
	\label{tab:PQAMorder4}
\end{table}

\subsection{Extended Feedback}\label{subsec:ExtFB}
 {We call \emph{Parity Symbol Generator} (PSG) the encoder block that 
computes the parity symbol sequences (see Figure~\ref{fig:New procedure}).}
Extended feedback consists of sending to the PSG a sequence of forward-channel
output observations over longer time intervals compared to  Deepcode~\cite{Kim2018}.
The PSG input column vector at the $k^{\rm th}$ iteration is defined
as follows:
\begin{equation}
	\label{eq:extInput}
	{\mathbf{i}}_k = \left[
	\begin{array}{c}
		x(k) \\
		\mathbf{n}_0(k-\delta_0:k) \\
		\mathbf{r}_0(k-\delta_1:k-1) \\
		\ldots \\
		\mathbf{r}_{P-1}(k-\delta_P:k-1) \\
	\end{array}
	\right],
\end{equation}
where $x(k)$ is the $k^{\rm th}$ systematic symbol, $\mathbf{n}_0(k-\delta_0:k)$ is 
a column vector of length $\delta_0+1$ which contains noise samples from the sequence $\mathbf{n}_0$ of~\eqref{eq:fwdChannel-Sys}, $\mathbf{r}_{l}(k-\delta_l:k-1)$ $(l=0,\ldots, P-1)$ is
a column vector of length $\delta_l$ which contains noise samples from the sequence
$\mathbf{r}_{l}$ of forward-channel noise samples that corrupt the $l^{\rm th }$ symbol of each parity
symbol sequence, that is:
\begin{equation}
	\mathbf{r}_{l} \triangleq (v_0(l),\ldots,v_{K-1}(l)),
\end{equation}
where $v_k(l)$ $(k=0,\ldots,K-1)$ is the $l^{\rm th}$ sample of $\mathbf{v}_{k}$ 
in~\eqref{eq:RXparitySeq} and $\delta_0, \ldots, \delta_P$ are arbitrary positive integers
($\delta_0$ can be 0), hereafter called the \emph{encoder input extensions}. 
We note that the Deepcode~\cite{Kim2018} encoder can be recovered as a
special case by setting $\delta_0 = 0$ and
$\delta_1 = \ldots = \delta_P = 1$, which means that, in each iteration,
	only a single noise sample for each systematic or parity check symbol is used.
The buffers in the DEF encoder contain the systematic symbol sequence
$\mathbf{x}$ and the corresponding forward-noise sequence $\mathbf{n}_0$
of \eqref{eq:fwdChannel-Sys}.
Those sequences are generated during the first encoding phase and used by
the PSG in the second phase.

\subsection{Parity Symbol Generator (PSG)}\label{subsec:PSG}
The core functionality of the DEF encoder is the computation of the
parity check symbols, which is performed by the block
denoted  (PSG) (see Fig.~\ref{fig:New procedure}).
PSG computes the $k^{\rm th}$ parity symbol sequence $\mathbf{p}_k$ based
on the $k^{\rm th}$ modulation symbol $x_k$ and a subset of the past
forward-channel outputs.

\begin{figure}[!t]
	\centering
	\includegraphics[scale=0.75,clip=true,trim=13.4cm 8.5cm 11.9cm 7.3cm]{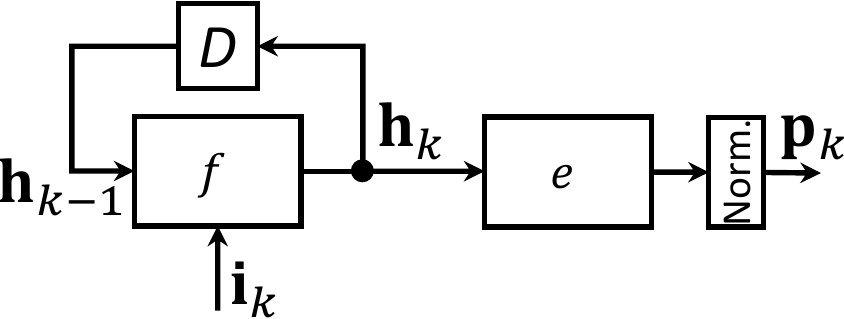}
	\caption{Structure of the PSG.}
	\label{fig:Parity symbol generator}
\end{figure}

Fig.~\ref{fig:Parity symbol generator} shows the structure of the PSG.
In the $k^{\rm th}$ encoding iteration, the PSG generates a $k^{\rm th}$ parity symbol sequence $\mathbf{p}_k$
which consists of $P$ real parity symbols obtained as follows:
\begin{equation}
\mathbf{p}_k = {\rm Norm}(e(\mathbf{h}_k)),
\label{Encoder_parity}
\end{equation}
where ${\bf{h}}_{k}$ -- a real vector of arbitrary length $H_0$ --
denotes the PSG state at time instant $k$, while function $e(\cdot)$ consists
of a linear transformation applied to the PSG state $\mathbf{h}_k$
obtained as follows:
\begin{equation}
	\label{eq:rate}
	e(\mathbf{h}_k) = \mathbf{A} \mathbf{h}_k + \mathbf{c},
\end{equation}
where $\mathbf{A}$ has size ${P\times H_0}$ and $\mathbf{c}$ has length ${P}$.
The above matrices $\mathbf{W}$, $\mathbf{Y}$, $\mathbf{A}$ and vectors $\mathbf{b}$,
$\mathbf{c}$ are obtained by NN training.
The ${\rm Norm}(\cdot)$ function normalizes the PSG output so that each parity symbol
has zero mean and unit variance.
The PSG state ${\bf{h}}_{k}$ is recursively computed as
\begin{equation}
\mathbf{h}_k = f(\mathbf{i}_k,\mathbf{h}_{k-1}),
\label{Encoder_state}
\end{equation}
where function $f(\cdot)$ will be discussed below, and
${\bf i}_k$ is defined in \eqref{eq:extInput}.
As for the initialization, we set $\mathbf{h}_0$ as
the all-zero vector.

Functions $e$ and $f$ will be parameterized using DNNs.
The structure of Fig.~\ref{fig:Parity symbol generator}
corresponds to a recurrent architecture, and therefore, we will 
consider the following three recurrent architectures to model it:
RNNs, GRUs and LSTMs.

\subsubsection{RNN}
When modelled with an RNN, the function $f(\cdot)$ in (\ref{Encoder_state})
is defined as follows:
\begin{equation}
f(\mathbf{i}_k,\mathbf{h}_{k-1}) = \tanh(\mathbf{W} \mathbf{h}_{k-1} + \mathbf{Y} \mathbf{i}_k + \mathbf{b}),
\label{eq:RNNstate}
\end{equation}
where $\mathbf{W}$ is a \emph{state-transition matrix} of size ${H_0\times H_0}$,
$\mathbf{Y}$ is an \emph{input-state matrix} of size ${H_0\times I}$ ($I$ is the length of vector $\mathbf{i}_k$),
 and $\mathbf{b}$ is a \emph{bias vector} of length ${H_0}$.
$\mathbf{W}$, $\mathbf{Y}$ and $\mathbf{b}$ are obtained by NN training.

\subsubsection{GRU}
With a GRU, the function $f(\cdot)$ of \eqref{Encoder_state}
	is defined as follows:
	\begin{eqnarray}
		f(\mathbf{i}_k,\mathbf{h}_{k-1}) & = & 
			f_0(\mathbf{i}_k,\mathbf{h}_{k-1}) \circ
			(1-z(\mathbf{i}_k,\mathbf{h}_{k-1})) \nonumber \\ 
			& + & \mathbf{h}_{k-1} \circ z(\mathbf{i}_k,\mathbf{h}_{k-1}).
		\label{eq:GRUstate}
	\end{eqnarray}
The function $f_0(\cdot)$ in \eqref{eq:GRUstate} is defined as follows:
\begin{eqnarray}
	f_0(\mathbf{i}_k,\mathbf{h}_{k-1}) & = & \tanh((\mathbf{W}_f \mathbf{h}_{k-1} + \mathbf{b}_h)
	\circ r(\mathbf{i}_k,\mathbf{h}_{k-1})\nonumber \\
	& + & \mathbf{Y}_f \mathbf{i}_k + \mathbf{b}_i).
		\label{eq:GRUstate2}
\end{eqnarray}
The functions $z(\cdot)$ in \eqref{eq:GRUstate} and $r(\cdot)$ in \eqref{eq:GRUstate2}
are defined as follows:
\begin{eqnarray}
	z(\mathbf{i}_k,\mathbf{h}_{k-1}) & = & \sigma(\mathbf{W}_z \mathbf{h}_{k-1} + \mathbf{Y}_z \mathbf{i}_k + \mathbf{b}_z) \\
	r(\mathbf{i}_k,\mathbf{h}_{k-1}) & = & \sigma(\mathbf{W}_r \mathbf{h}_{k-1} + \mathbf{Y}_r \mathbf{i}_k + \mathbf{b}_r)
		\label{eq:GRUstate3}
\end{eqnarray}
where $\sigma(x)\triangleq (1+e^{-x})^{-1}$ denotes the 
\emph{sigmoid} function. In eqns. \eqref{eq:GRUstate}-\eqref{eq:GRUstate3}, matrices $\mathbf{W}_f, \mathbf{W}_z, \mathbf{W}_r, \mathbf{Y}_f, \mathbf{Y}_z, \mathbf{Y}_r$ and vectors $\mathbf{b}_h, \mathbf{b}_i, \mathbf{b}_z,\mathbf{b}_r$ are obtained by NN training.

\subsubsection{LSTM}
As for LSTM, the function $f(\cdot)$ of \eqref{Encoder_state}
is defined as follows:
\begin{eqnarray}
	f(\mathbf{i}_k,\mathbf{h}_{k-1}) & = & 
	f_1(\mathbf{i}_k,\mathbf{h}_{k-1}) \circ
	\tanh(\mathbf{s}_k)
	\label{eq:LSTMstate}
\end{eqnarray}
where $\mathbf{s}_k$ is the \emph{cell state} at time instant $k$. The cell state provides long-term memory capability to the LSTM NN, whereas the state $\mathbf{h}_k$ provides short-term memory capability.
The cell state is recursively computed as follows:
\begin{eqnarray}
	\mathbf{s}_k & = & f_2(\mathbf{i}_k,\mathbf{h}_{k-1}) \circ \mathbf{s}_{k-1} \nonumber \\ 
	& + & f_3(\mathbf{i}_k,\mathbf{h}_{k-1}) \circ f_4(\mathbf{i}_k,\mathbf{h}_{k-1}). \label{eq:LSTM2}
\end{eqnarray}
The function $f_1$ in \eqref{eq:LSTMstate} and functions $f_2, f_3$ and $f_4$
in  \eqref{eq:LSTM2} are defined as follows:
\begin{eqnarray}
	f_1(\mathbf{i}_k,\mathbf{h}_{k-1})\!&\!=\!& \!\sigma(\mathbf{W}_1 \mathbf{h}_{k-1} + \mathbf{Y}_1 \mathbf{i}_k + \mathbf{b}_1) \label{eq:LSTMstate0} \\
	f_2(\mathbf{i}_k,\mathbf{h}_{k-1})\!&\!=\!& \!\sigma(\mathbf{W}_2 \mathbf{h}_{k-1} + \mathbf{Y}_2 \mathbf{i}_k + \mathbf{b}_2) \\
	f_3(\mathbf{i}_k,\mathbf{h}_{k-1})\!&\!=\!& \!\sigma(\mathbf{W}_3 \mathbf{h}_{k-1} + \mathbf{Y}_3 \mathbf{i}_k + \mathbf{b}_3) \\
	f_4(\mathbf{i}_k,\mathbf{h}_{k-1})\!&\!=\!& 
	\!\tanh(\mathbf{W}_4 \mathbf{h}_{k-1} + \mathbf{Y}_4 \mathbf{i}_k + \mathbf{b}_4) \label{eq:LSTMstate3}
\end{eqnarray}
In eqns. \eqref{eq:LSTMstate0}-\eqref{eq:LSTMstate3}, matrices $\mathbf{W}_1$, $\mathbf{W}_2$, $\mathbf{W}_3$, $\mathbf{W}_4$, $\mathbf{Y}_1$, $\mathbf{Y}_2$, $\mathbf{Y}_3$, $\mathbf{Y}_4$ and vectors $\mathbf{b}_1$, $\mathbf{b}_2$, $\mathbf{b}_3$, $\mathbf{b}_4$ are obtained by NN training.

\subsection{Mitigation of Unequal Bit Error Distribution}
It has been observed in~\cite{Kim2018} that the feedback codes based
on RNNs exhibit a non-uniform bit error distribution,
i.e., the final message bits typically have a
significantly larger error rate compared to other bits.
In order to mitigate the detrimental effect of non-uniform bit
error distribution, ~\cite{Kim2018} introduced two countermeasures:
\begin{itemize}
	\item \emph{Zero-padding}. Zero-padding consists in appending at least
	one information bit with pre-defined value (e.g., zero) at the
	end of the message. The appended information bit(s) are discarded
	at the decoder, such that the positions 
	affected by higher error rates carry no information.
	\item \emph{Power reallocation}. Zero-padding alone is not enough
	to mitigate unequal errors, and moreover it reduces the effective code rate.
	Instead, power reallocation redistributes
	the power among the codeword symbols so as to provide better error
	protection to the message bits whose positions are more error-prone,
	i.e., the initial and final positions.
\end{itemize}

\begin{figure}[!b]
	\centering
	\includegraphics[scale=0.75,clip=true,trim=12.9cm 6.3cm 11.6cm 7.2cm]{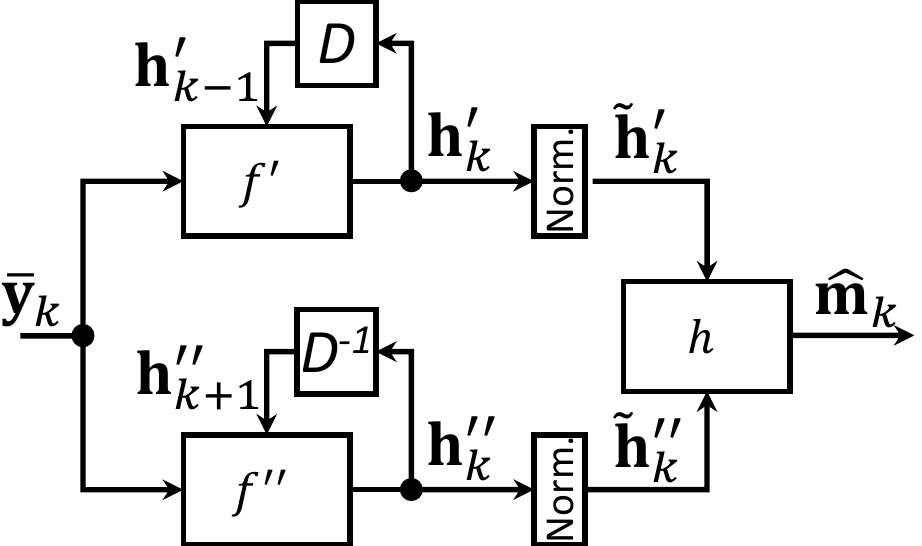}
	\caption{DEF decoder.}
	\label{fig:DEFdec}
\end{figure}

\subsection{DEF Decoder}\label{Rx}
In DNN-based codes, encoder and decoder are implemented as two
separate DNNs whose coefficients are determined through a joint
encoder-decoder training procedure.
Therefore, the \emph{encoder} structure has impact on the \emph{decoder}
coefficients obtained through training, and vice-versa.
In that sense, the chosen \emph{decoder} structure has impact on the
resulting code.

The DEF decoder maps the received DEF codeword to a decoded message
$\mathbf{\hat{m}}$ as follows:
\begin{equation}
\mathbf{\hat{m}} = g(\mathbf{\bar{x}}, \mathbf{\bar{p}}^{(1)},\ldots,\mathbf{\bar{p}}^{(K)}).
\end{equation}
The decoder consists of a bi-directional {recurrent NN -- GRU or LSTM --} followed by
a linear transformation {and a sigmoid function}.
The bi-directional recurrent NN computes a sequence of
forward-states $\mathbf{h}_k'$ and backward-states $\mathbf{h}_k''$
as follows:
\begin{eqnarray}
\mathbf{h}_k' & = & f'(\mathbf{\bar y}_k,\mathbf{h}_{k-1}') \\
\mathbf{h}_{k-1}'' & = & f''(\mathbf{\bar y}_k,\mathbf{h}_{k}'')
\label{Decoder_state}
\end{eqnarray}
where functions $f', f''$ are defined as in \eqref{eq:GRUstate}
{for the GRU-based decoder and as in \eqref{eq:LSTMstate} for
the LSTM-based decoder,} and
the input column vector $\mathbf{\bar y}_k$ is defined as follows:
\begin{equation}
\mathbf{\bar y}_{k} = \left[
\begin{array}{c}
\mathbf{\bar x}(k-\gamma_0:k) \\
\mathbf{\bar q}_0(k-\gamma_1:k) \\
\ldots\\
\mathbf{\bar q}_{P-1}(k-\gamma_P:k)
\end{array}
\right],
\label{DecoderInput}
\end{equation}
where $\mathbf{\bar x}(k-\gamma_0:k)$ is a column vector of length $\gamma_0+1$
	which contains symbols from the received systematic sequence $\mathbf{\bar x}$ of~\eqref{eq:fwdChannel-Sys},
	and $\mathbf{\bar q}_l(k-\gamma_l:k), l=0,\ldots,P-1$,
	is a column vector of length $\gamma_l+1$ containing symbols from the sequence
	$\mathbf{\bar q}_{l}$, which consists of the $l^{\rm th}$ symbol of each received parity	sequence $\bar{\mathbf{p}}_{k}$~\eqref{eq:RXparitySeq}.
	$\mathbf{\bar q}_{l}$ is defined as follows:
\begin{equation}
	\mathbf{\bar q}_{l} \triangleq ({\bar p}_0(l),\ldots,{\bar p}_{K-1}(l)),l=0,\ldots,P-1.
\end{equation}
Finally, the values $\gamma_0, \ldots,\gamma_P$ are arbitrary non-negative
integers, hereafter called the \emph{decoder input extensions}.
The initial forward NN state $\mathbf{h}_0'$ and the initial backward NN state
$\mathbf{h}_K''$ are set as all-zero vectors.

The $k^{\rm th}$ decoder output is obtained as follows:
\begin{equation}
\hat{\mathbf{m}}_k = h({\mathbf{\tilde h}_k', \mathbf{\tilde h}_{k-1}''}) \triangleq \sigma\left(\mathbf{C} \left[\begin{array}{l}
\mathbf{\tilde h}_k' \\
\mathbf{\tilde h}_{k-1}''
 \end{array}\right] + \mathbf{d}\right),
\end{equation}
where $\sigma(\cdot)$ is the \emph{sigmoid} function,
$\mathbf{C}$ is a matrix of size $Q/2 \times 2 H_0$, and $\mathbf{d}$ is a vector of size $Q/2$.
$\mathbf{C}$ and $\mathbf{d}$ are obtained by NN training.
{Vectors $\mathbf{\tilde h}_k'$ and $\mathbf{\tilde h}_k''$ are obtained by 
	normalizing vectors $\mathbf{h}_k'$ and $\mathbf{h}_k''$
	so that each element of $\mathbf{\tilde h}_k'$ and $\mathbf{\tilde h}_k''$ has zero
	mean and unit variance.}
Vector $\mathbf{\hat m}_k$ provides the estimates of the message bits in a corresponding $Q/2$-tuple,
that is:
\begin{equation}
\mathbf{\hat m}_k = (\hat{m}(kQ/2), \ldots, \hat{m}((k+1)Q/2-1)).
\end{equation}

The Deepcode decoder from~\cite{Kim2018} is recovered by setting
$\gamma_l = 0,l=0,1,...,P$ in \eqref{DecoderInput}.

\section{Transceiver Training}\label{sec:training}
The coding and modulation schemes used in conventional
	communication systems are optimized for a given SNR range.
We take the same approach for DNN-based codes --
as DNN code training produces different codes depending on the training
SNR, we divide the target range of forward SNRs into (small) non-overlapping intervals
and select a single training SNR within each interval.

Encoder and decoder are implemented as two separate DNNs whose coefficients
are determined through a joint training procedure.
The training procedure consists in the transmission of batches
of randomly generated messages.
The number of batches is $2 \times 10^4$, where each {batch}
contains $2\times 10^3$ messages.
DNN coefficients are updated by an ADAptive Moment (ADAM) estimation
	optimizer based on the \emph{binary cross-entropy} (BCE)
		loss function.
		For each batch, a loss value is obtained by computing the BCE
		between the messages in that batch and the corresponding
		decoder outputs.
The learning rate is initially set to $0.02$
and divided by $10$ after the first group of $10^3$ batches.
The gradient magnitude is clipped to $1$.

{By monitoring the BCE loss value
	throughout the entire training session, we noticed that the loss
	trajectory has high peaks which appear more frequently during
	the initial phases of training.
	Those peaks indicate that the training process is driving the 
	encoder/decoder NNs away from their optimal performances.
In order to mitigate the detrimental effect of the above events,
{the following countermeasures have been taken:
\begin{itemize}
	\item usage of a larger {batch} size -- 10 times larger than~\cite{Kim2018}.
		Usage of large {batches} stabilizes training\footnote{By 
			\emph{training stabilization} we mean that the loss function produces smoother
			trajectories during training.} and accelerates convergence of NN weights towards
		values that produce good performance;
	\item implementation of a training \emph{roll-back} mechanism that discards the
	NN weight updates of the last epoch if the loss value produced
	by the NNs with updated weights is at least 10 times larger than
	the loss produced by the NNs with previous weights.
\end{itemize}}

{As we observed that the outcome of training is sensitive to the
	random number generators' initialization,} each training is repeated three
times with different initialization seeds.
For each repetition, we record the final NN weights and the NN
weights that produced the smallest loss during training.
After training, link-level simulations (LLS) are performed using 
all the recorded weights. The set of weights that provides the
lowest block error rate (BLER) is kept and the others are discarded.
\begin{table}[!b]
	\centering
	\renewcommand{\arraystretch}{1.3}
	\begin{tabular}{c|c}
		\bf Training parameter  & \bf Values \\
		\hline \hline
		Number of epochs &  2000 \\ \hline
		Number of batches per epoch & 10 \\ \hline
		Number of codewords per batch & 2000 \\ \hline
		Training message length [bits] & 50 \\ \hline
		Starting epoch for codeword-level weights training & 100 \\ \hline
		Starting epoch for symbol-level weights training & 200 \\ \hline
	\end{tabular}
	\caption{Training parameters.}
	\label{tab:TrainParam}
\end{table}

As described in Subsection \ref{subsec:PSG} and illustrated in
Fig.~\ref{fig:Parity symbol generator},
the PSG output is normalized so that each coded symbol has zero
mean and unit variance.
During NN training, normalization subtracts the \emph{batch mean} from the
PSG output and divides the result of subtraction by the batch standard deviation.
After training, \emph{encoder calibration} is performed in order to compute the mean and
the variance of the RNN outputs over a given number of codewords. 
Calibration is done over $10^6$ codewords in the simulations here reported.
In LLS, normalization is done using the mean and variance
values computed during calibration.

The training strategy for the encoder's codeword and symbol power levels has
been optimized empirically.
The levels are initialized to unit value, and kept constant for
a given number of epochs as early start of training produces codes with poor performance.
On the other hand, if training of levels is started too late, they
remain close to their initial unit value, and therefore produce no benefits.
It has been found empirically that starting to train codeword power levels at epoch
100 and symbol power levels at epoch 200 provides the best results.

As suggested in~\cite{Kim2018}, it may be beneficial to perform training with longer messages
compared to link level evaluation as training with short messages does not produce good codes.
According to our observations, training with longer messages -- twice the length of
LLS messages -- 
is beneficial. However, according to our observations, the benefit of using
longer messages vanishes when training with larger batches.
Therefore, in our evaluations the length of training messages and LLS messages
is the same.

The above training method produces codes with better performance
compared to the method of~\cite{Kim2018}, as the
performance evaluations of Section~\ref{Perf} will show.
{Training parameters are summarized in Table~\ref{tab:TrainParam}.}
}

\begin{figure}[!t]
	\centering
	\resizebox{1.0\hsize}{!}{
		\includegraphics[clip=true,trim=2.1cm 7.7cm 2.3cm 9.4cm]{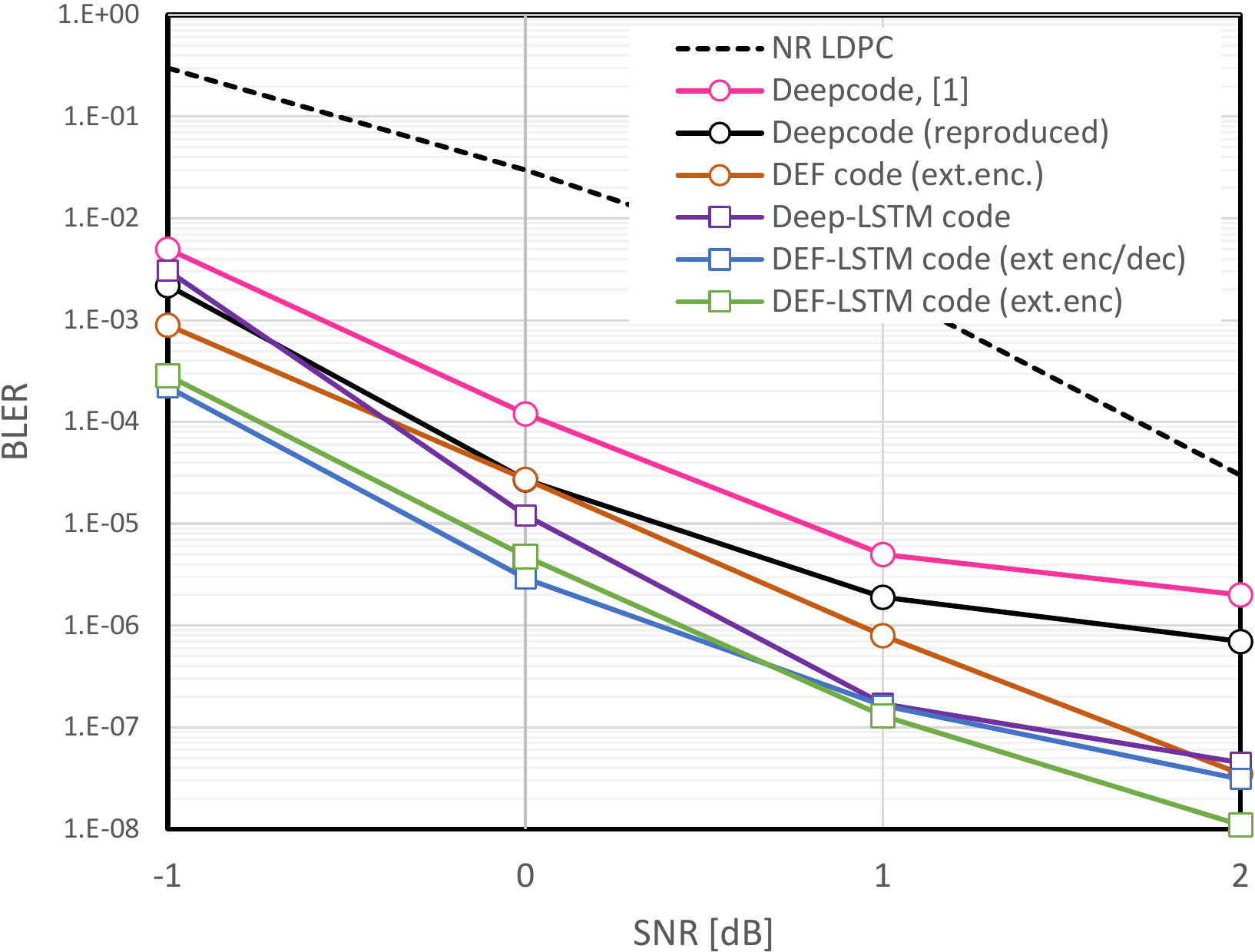}}
	\caption{Performance comparison of Deepcode, {DEF codes,
			LSTM-based Deepcode, and {DEF-LSTM} codes.
			Spectral efficiency is 0.67 bits/s/Hz} ($Q=2$, $P=2$).}
	\label{fig:BLERvsSNR}
\end{figure}

\section{Performance Evaluations}\label{Perf}
{In this section, we assess the BLER performance of DEF
	codes and compare their performance with the performance of the NR LDPC code reported in~\cite{bib:R1-1713740} and the performance of Deepcode~\cite{Kim2018} for the same
	spectral efficiency (SE).
The SE is defined as the ratio of the number of information bits $L$
over the number of forward-channel time-frequency resources used
for transmission of the corresponding codeword}.
As each time-frequency resource carries a complex symbol, and since each
complex symbol is produced by combining two consecutive real symbols, we have
\begin{equation}
SE  \triangleq \frac{Q}{1+P} \;\rm [bits/s/Hz].
\end{equation}
The forward-channel and feedback-channel impairments are modeled as
AWGN with variance $\sigma_n^2=1/{SNR}$ and
$\sigma_{FB}^2=1/{SNR_{FB}}$, respectively.
{The training forward SNR and LLS forward SNR are the same;
	the feedback channel is noiseless.}

The set of parameters used in the performance
evaluations is shown in Table~\ref{tab:EvalParam}.
For DEF code performance evaluations, we show that even the
shortest feedback extensions -- corresponding to the $\delta$ and $\gamma$
parameters of Table~\ref{tab:EvalParam} -- produces significant gains.
The investigation of performance with larger feedback extensions is left
for future work.
Details of the evaluated architectures are reported in Table~\ref{tab:codeArch}.

\begin{figure}[!t]
	\centering
	\resizebox{1.0\hsize}{!}{
		\includegraphics[clip=true,trim=2.1cm 7.7cm 2.1cm 9.3cm]{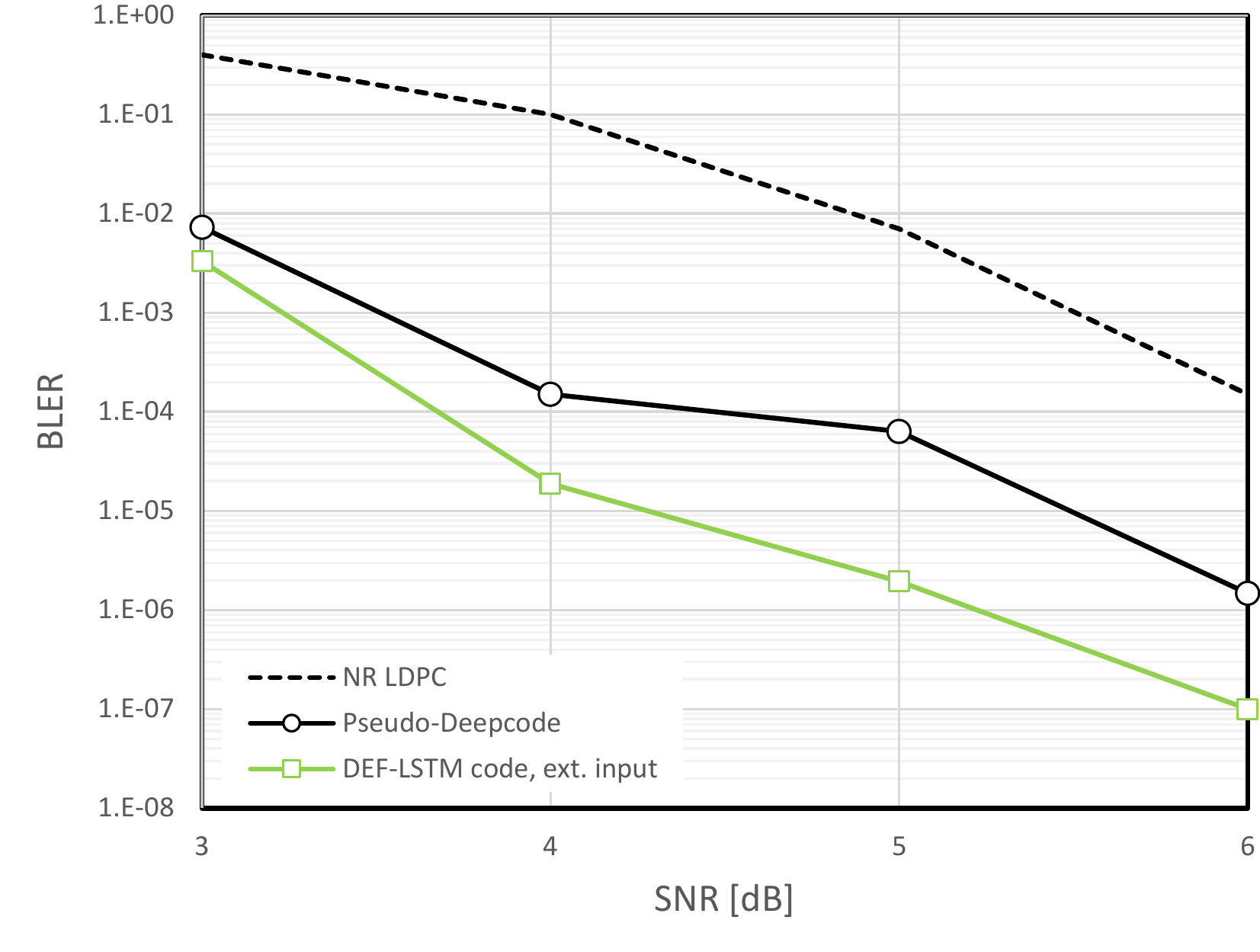}}
	\caption{Performance comparison of Deepcode, pseudo-Deepcode,
		and {DEF-LSTM} code with extended encoder input.
		Spectral efficiency is 1.33 bits/s/Hz ($Q=4$, $P=2$).}
	\label{fig:BLERvsSNR1}
\end{figure}

\begin{table}[!b]
\centering
\renewcommand{\arraystretch}{1.3}
\begin{tabular}{c|c}
\bf DEF code parameter  & \bf Selected values \\
\hline \hline
$K$ [symbols] & 50 \\ \hline
$P$  & 2 \\ \hline
$H_0$ & 50 \\ \hline
\# zero-padding bits & 1 \\ \hline
Encoder input extensions & $(\delta_0,\delta_1,\delta_2)=(1,2,2)$ \\ \hline
Decoder input extensions & $(\gamma_0,\gamma_1,\gamma_2) = (1,1,1) $
\end{tabular}
\caption{Evaluation parameters.}
\label{tab:EvalParam}
\end{table}

\begin{table}[!b]
		\centering
		\renewcommand{\arraystretch}{1.3}
		\begin{tabular}{c|c|c}
			\bf Code & \bf Encoder NN & \bf Decoder NN \\
			& \bf (type, \#layers) & \bf (type, \#layers) \\
			\hline \hline
			Deepcode & RNN, 1 & bidir. GRU, 2 \\ \hline
			DEF code & RNN, 1 & bidir. GRU, 2 \\ \hline
			Deep-LSTM code & LSTM, 1 & bidir. LSTM, 2 \\ \hline
			DEF-LSTM code & LSTM, 1 & bidir. LSTM, 2 \\ \hline
		\end{tabular}
		\caption{Evaluated architectures.}
		\label{tab:codeArch}
	\end{table}

Fig.~\ref{fig:BLERvsSNR} shows the Block Error Rate
(BLER) vs. forward SNR of several codes with $SE=0.67$ bits/s/Hz. 
The plot shows Deepcode~\cite{Kim2018} (pink curve), 
Deepcode obtained by the training method of Sec.~\ref{sec:training}
(solid black curve), { DEF code with extended encoder input (orange curve),}
Deepcode with LSTM-based encoder and decoder NNs
(purple curve), DEF code with extended encoder input (green curve) and 
DEF code with extended encoder and decoder input (blue curve).
All DNN-based codes use second-order modulation (i.e., $Q=2$) and $P=2$
parity symbols per systematic symbol.
Thus, the corresponding SE is $0.67$ bits/s/Hz.
The performance of the NR LDPC code as reported in~\cite{bib:R1-1713740}
with the same SE (QPSK modulation, code rate 1/3) is shown by
a dashed black curve.

Based on the data shown in Fig.~\ref{fig:BLERvsSNR}, the following
observations are made:
\begin{itemize}
	\item {The DEF code with extended encoder input (orange curve) has better
		performance than Deepcode (solid black curve);}
	\item The {{DEF-LSTM} codes (green and blue curves) have} the
	best performance among all the evaluated codes;
	\item  {DEF-LSTM} code with extended encoder input and {DEF-LSTM}
	code with extended encoder/decoder input have similar performance except for high SNRs, 
	where the former performs slightly better;
	\item {DEF-LSTM} codes (green and blue curve) outperform NR LDPC
	(dashed black curve) by at least three orders of magnitude BLER	for all SNRs.
	\item The training method of Section~\ref{sec:training} (black curve)
produces codes with better performance than the training method of~\cite{Kim2018} (pink curve).  
\end{itemize}
{
Based on the first observation above, it can be concluded that \emph{encoder}
input extension produces performance improvements.
{Subsequent observations highlight that the encoder input extension provides
performance improvements when combined with LSTM.
However, based on the observation in the third bullet, we can conclude that \emph{decoder} input
	extension brings no benefits compared to \emph{encoder} input extension.} 
Moreover, the above performance evaluations show that usage of LSTM
in the encoder and decoder provides significant performance
improvements compared to RNN/GRU based codes.}
 
Figure~\ref{fig:BLERvsSNR1} shows the BLER performance of DNN-based
	codes with modulation order $Q=4$ -- the corresponding SE is $1.33$ bits/s/Hz.
As Deepcode~\cite{Kim2018} is not defined for  SEs higher than
	$0.67$ bits/s/Hz, we implemented a \emph{pseudo-Deepcode} by
	replacing the Deepcode modulator with a modulator of order $Q=4$.
Results show that the {DEF-LSTM} code has better performance compared to the
pseudo-Deepcode as its BLER is significantly lower in the whole range of SNR
that we evaluated.
The {DEF-LSTM} code BLER gain over {pseudo-}Deepcode is larger than one order of magnitude for
SNR=5 dB and 6 dB.
Moreover, the {DEF-LSTM} code outperforms NR LDPC (dashed black curve) by at
least three orders of magnitude BLER for SNR $\geq4$ dB.

\section{Conclusion and further work}\label{conclude}

A new {deep-neural-network based error correction
 encoder architecture} for channels
with feedback has been presented.
It has been shown that the codes designed according to the {DEF} architecture
achieve {lower error rates than any other code designed for channels with feedback.
Moreover, by a suitable selection of the modulation order, these codes can adapt to the forward
channel quality, thereby providing the maximum spectral efficiency that is attainable
for the given forward channel quality.

Study of these codes in more realistic scenarios such as noisy feedback channels and fading is
left for future work.}

\end{document}